# Deep Learning based Spectral CT Imaging

Weiwen Wu[1,4], Dianlin Hu[2], Chuang Niu[4], Lieza Vanden Broeke[1], Anthony P.H. Butler[3], Peng Cao[1], James Atlas[3], Alexander Chernoglazov[3], Varut Vardhanabhuti[1,*], Ge Wang[4]

[1]Dept of Diagnostic Radiology, Queen Mary Hospital, University of Hong Kong, Hong Kong, People's Republic of China

[2]The Laboratory of Image Science and Technology, Southeast University, Nanjing, People's Republic of China

[3]Department of Radiology, University of Otago, Christchurch, New Zealand

[4]Biomedical Imaging Center, Center for Biotechnology and Interdisciplinary Studies, Department of Biomedical Engineering, School of Engineering, Rensselaer Polytechnic Institute, Troy, New York, USA

Correspondence author: V. Vardhanabhuti (varv@hku.hk)

**Abstract:** Spectral computed tomography (CT) has attracted much attention in radiation dose reduction, metal artifacts removal, tissue quantification and material discrimination. The x-ray energy spectrum is divided into several bins, each energy-bin-specific projection has a low signal-noise-ratio (SNR) than the current-integrating counterpart, which makes image reconstruction a unique challenge. Traditional wisdom is to use prior knowledge based iterative methods. However, this kind of methods demands a great computational cost. Inspired by deep learning, here we first develop a deep learning based reconstruction method; i.e., **U**-net with $L_p^p$-norm, **T**otal variation, **R**esidual learning, and **A**nisotropic adaption (ULTRA). Specifically, we emphasize the various multi-scale feature fusion and multichannel filtering enhancement with a denser connection encoding architecture for residual learning and feature fusion. To address the image deblurring problem associated with the $L_2^2$- loss, we propose a general $L_p^p$-loss, $p > 0$. Furthermore, the images from different energy bins share similar structures of the same object, the regularization characterizing correlations of different energy bins is incorporated into the $L_p^p$- loss function, which helps unify the deep learning based methods with traditional compressed sensing based methods. Finally, the anisotropically weighted total variation is employed to characterize the sparsity in the spatial-spectral domain to regularize the proposed network. In particular, we validate our ULTRA networks on three large-scale spectral CT datasets, and obtain excellent results relative to the competing algorithms. In conclusion, our quantitative and qualitative results in numerical simulation and preclinical experiments demonstrate that our proposed approach is accurate, efficient and robust for high-quality spectral CT image reconstruction.

## I. Introduction

The spectral CT has huge potentials in aspects of tissue characterization (Boll, et al., 2009), radiation dose reduction, material discrimination, x-ray beam hardening artifacts reduction, and quantitative analysis for tissue component (Long & Fessler, 2014). The spectral CT system is usually equipped with the photon-counting detector (PCD) to provide multiple quantified materials by measuring each x-ray narrow energy bin attenuations. This means the number of received photons within one narrow energy bin is small, which can cause noise improvement within the projection. Besides, due to the limitations of PCD manufacturing, consistent detector responses are difficult to achieve and lead to data inconsistency between measurements and real case. These factors can cause the artefacts in spectral CT images. PCD can discriminate x-ray photon energy by adjusting energy thresholds. However, x-ray radiation, material component distribution and system geometrical configuration play an important role in threshold selection. Besides, different PCD detector units have different responses to individual x-ray photon energy. This can further result in spectral distortions, including charge sharing, K-escape, fluorescence x-ray emission and pulse pileups. All these factors make data inconsistency between measurements and theoretical values. This results in severe measurement noise in projections and it further compromises the significance of spectral CT in practice (Taguchi, Stierstorfer, Polster, Lee, & Kappler, 2018). How to achieve satisfactory results from such noisy measurements currently is a major challenge for spectral CT applications. To address this issue, many advanced spectral CT iteration image reconstruction models were proposed.

Since there is severe noise within projections, it is difficult to apply the traditional analytic methods directly, such as filtered back projection (FBP) method for spectral CT. To obtain high-quality spectral CT images, it appears feasible to extend the traditional CT iteration image reconstruction methods to spectral CT, such as total variation (TV) (Xu, et al., 2012), dual-dictionary learning (Zhao, et al., 2012), piecewise linear tight frame transform (Zhao, Gao, Ding, & Molloi, 2013), *etc*. However, these methods only consider image features from channel-independent and ignore the correlation between different energy bins. Besides, the selection of parameters depending on the number of energy bins will degrade its practicality. Introducing high-quality image into the reconstruction model is feasible, including total image constrained diffusion tensor (Niu, et al., 2019) and spectral prior image constrained compressed sensing (Yu, Leng, Li, & McCollough, 2016). However, obtaining such high-quality prior images in practice is not always possible. Considering the sparsity or low-rank properties within spatial-spectral domain, tensor-based nuclear norm (Semerci, Hao, Kilmer, & Miller, 2014), prior rank, intensity and sparsity model (Chu, Li, Chen, Wang, & Gao, 2012; Gao, Yu, Osher, & Wang, 2011), total nuclear variation (Rigie & Rivière, 2015), patch-based low-rank (Kim, et al., 2015), structure tensor TV (Zeng, et al., 2016), tensor dictionary learning (Chen, Zhang, Kalra, et al., 2017; Wu, Zhang, Wang, Liu, Chen, et al., 2018) are developed. To further improve the reconstructed image quality, the non-local image similarity was explored in spectral CT reconstruction, including spatial spectral nonlocal means (B. Li, et al., 2018), nonlocal low-rank and sparse matrix decomposition (Niu, Yu, Ma, & Wang, 2018), spatial-spectral non-local means(Z. Li, Leng, Yu, Manduca, & McCollough, 2017), nonlocal spectral similarity(Yao, et al., 2019), aided by self-similarity in image-spectral tensors (Xia, et al., 2019), spatial-spectral cube matching frame (Wu, Zhang, Wang, Liu, Wang, et al., 2018), non-local low-rank cube-based tensor factorization (Wu, Liu, Zhang, Wang, & Yu, 2019), weight adaptive total-variation and image-spectral tensor factorization (Wu, Hu, An, Wang, & Luo, 2020), spectral-Image decomposition with energy-fusion sensing (S. Wang, et al., 2021), $L_0$-norm based adaptive SPICCS (S. Wang, et al., 2021), *etc*. All of these methods are proposed for fan-beam geometry spectral CT. However, it remains challenging in practice due to its large computation costs, especially for cone-beam geometry. Besides, these methods usually contain several tunable parameters which need to be chosen for specific clinical applications. Appropriate parameters selection remains a problem. Finally, the image structures and details are still blurred, especially in preclinical applications.

Deep learning (DL) as an advanced signal processing technique has been attracting great attention for different computer vision tasks (LeCun, Bengio, & Hinton, 2015). Regarding the CT imaging groups, DL as an emerging technique has proven its potentials in CT imaging (Bao, et al., 2019; Chen, et al., 2018; Ge, et al., 2019; Huang, et al., 2020; Lyu, et al., 2021), including sparse-view reconstruction (Chen, Zhang, Kalra, et al., 2017; Chen, Zhang, Zhang, et al., 2017; Wu, et al., 2021), limited-angle CT (J. Wang, Zeng, Wang, & Guo, 2019), artifacts removal (Y. Zhang & Yu, 2018), dose reduction (Ahn, Heo, & Kim, 2019; Cong, et al., 2019), interior CT (Han & Ye, 2019), scatter correction (Maier, Sawall, Knaup, & Kachelrieß, 2018) and so on. To extend its applications to spectral CT, deep learning was employed to material decomposition (Lu, et al., 2019) and ring artifact reduction (Fang, Li, & Chen, 2020). However, as far as the authors are aware, there is no deep learning based techniques employed for spectral CT imaging. There are two issues that need to be addressed with consideration of deep learning based spectral CT reconstruction. First, how to obtain adequate ground truth (label) and noisy datasets (training datasets). Second, how to design advanced deep learning network to obtain good reconstruction performance. To overcome these issues, we propose a deep learning based reconstruction method, i.e., **U**-net with $L_p^p$-norm, **T**otal variation, **R**esidual learning, and **A**nisotropic adaption (ULTRA). The contribution of this study can be summarized into four points:

First, we design a database for spectral CT based on conventional CT images. Specifically, the conventional CT images can be segmented into materials independently by a threshold method. According to Beer's law, we can generate the measurement of one x-ray by the multiplication of specified material and linear attenuation coefficients. Note that the shape of x-ray energy and the corresponding thresholds of each energy bin. The Poisson noise is added to all spectral CT measurements. Finally, the conventional filtered back projection (FBP) is employed to reconstruct noisy-free spectral CT images and noisy spectral CT images respectively. These noisy-free and noisy images are ideal label and training dataset.

Second, the U-net network is a classical network for image segmentation, and sparse view reconstruction for traditional CT. In this study, it is employed for spectral CT reconstruction. To avoid the image feature missing during the encoding process of U-net, we design a skip-encode U-net, in which the different size of feature maps can fully communicate by a concatenating strategy to extract more image features, which can further improve the performance of the network.

Third, the loss function of conventional network U-net network is mainly based on $L_2^2$- loss, which can cause image details missing and image edge blurred. In this study, we propose a general $L_p^p(p>0)$- to overcome this issue. In this case, we can adjust the value of $p$ to obtain better performance. Besides, we derive the sub-gradient of $L_p^p$ and clarify the stability of this loss function.

Fourth, imaged object based prior is usually incorporated into iteration based spectral CT reconstruction to obtain better results. In fact, this type of methods considers the property of final results. As deep learning based CT reconstruction usually ignore such prior, we first fuse such prior into loss function and formulate a general prior knowledge driven loss function. To further clarify this point, the anisotropically weighted total variation containing the correlation of spectral information is considered to characterize the sparsity property of network output.

The rest of this work is organized as follows. In Section II, we introduce the related work for spectral CT reconstruction and then design a spectral CT database. In Section III, we introduce the proposed network and the proposed prior knowledge driven loss function. In Section IV, the experimental design is presented, and results are described and analyzed. Finally, in Section V we discuss relevant problems and make our conclusion.

## II. Related Work

**A. Noise reduction for Spectral CT**

Regarding the reconstruction methods for spectral CT, they can be divided into the following three classes, i.e., sinogram filtration (Noh, Fessler, & Kinahan, 2009), regularization-based iteration reconstruction(Wu, Zhang, Wang, Liu, Chen, et al., 2018) and post-processing technique (Y. Zhang, Salehjahromi, & Yu, 2019). Sinogram filtration-based algorithms are mainly focusing on performing denoising operator on either raw photon or log-transformed datasets before image reconstruction. Different from conventional CT, the noise from PCD is complicated, which result in the difficulty in designing the sonogram filter by using a straight-forward manner. Since back projection operator of the image reconstruction methods can magnify noise within measurements, those methods usually cause edge blurring or resolution loss.

Regularization-based iterative reconstruction methods have been attracting great attention over the past decade, especially in the field of spectral CT. To further clarify this point, the imaging geometry of spectral CT with fan-beam geometry can be formulated as follow

$$\mathbf{A}\boldsymbol{\mu}_m = \mathbf{p}_m + \boldsymbol{\epsilon}_m, m=1,\dots,M \tag{1}$$

$\mathbf{A} \in \mathcal{R}^{L \times N_1}$ represents the forward projection matrix and L represents the number of x-rays. $N_1 = N_w \times N_h$ is the pixel number of $\boldsymbol{\mu}_m \in \mathcal{R}^{N_1 \times 1}$, where $N_w$ and $N_h$ are the width and height of the image from $m^{th}$ energy bin. $\mathbf{p}_m \in \mathcal{R}^{L \times 1}$ and $\boldsymbol{\epsilon}_m \in \mathcal{R}^{L \times 1}$ are the projection and noise of $m^{th}$ energy bin. Introducing the prior knowledge into the imaging model, the general formula can be written as

$$\min_{\mathcal{U}} \left( \frac{1}{2} \sum_{m=1}^{M} \|\mathbf{A}\boldsymbol{\mu}_m - \mathbf{p}_m\|_F^2 + \lambda \mathcal{F}(\mathcal{U}) \right), \tag{2}$$

where $\mathcal{U}$ is the tensor format of $\{\boldsymbol{\mu}_m\}_{m=1}^{M}$, $\lambda$ is regularization parameters and $\mathcal{F}$ represents a specified domain transform. Specifically, different transforms $\mathcal{F}$ can lead to different reconstruction model. Commonly-used prior knowledge includes non-negativity, total variation (Xu, et al., 2012), tensor dictionary learning [18], tensor factorization[19], etc. To obtain the solution, the optimized strategy is usually employed to obtain the solution (Y. Zhang, Mou, Wang, & Yu, 2017). As for this kind of methods, the prior knowledge selections play an important role in final results, this means different applications may correspond to different regularization priors. Besides, the parameters selections may also become a challenge in practice. Finally, because this kind of mathematical model need to be optimized by alternative iteration, these techniques are time-consuming, especially for cone-beam geometry.

Different to the sonogram filtration and regularization-based iteration methods, the post-processing mainly focus on performing denoising techniques on the reconstructed spectral CT image and its general mathematical model can be given as

$$\min_{\mathcal{V}} \left( \frac{1}{2} \|\mathcal{V} - \mathcal{U}\|_F^2 + \lambda \mathcal{F}(\mathcal{V}) \right). \tag{3}$$

Here $\mathcal{U}$ is reconstructed by FBP, ART or SART. Traditional spectral CT post-processing methods including non-local means (Pan, Liu, de Ruiter, & Grasset, 2009), tensor decomposition (Y. Zhang, et al., 2019), block-matching 4D (Wu, Zhang, Wang, Liu, Wang, et al., 2018) and so on. However, these techniques may result in uneven performance improvements, potential over-smoothing, and critical subtle structural details being obscured.

With the development of deep learning techniques, they are rapidly employed in many aspects of conventional CT, including low-dose cases (Chen, Zhang, Kalra, et al., 2017; Yang, et al., 2018), limited-angle CT(Bubba, et al., 2019; Würfl, et al., 2018), interior problem(Han & Ye, 2019), metal artifacts reduction (Y. Zhang & Yu, 2018), and ring artifacts removal (Chang, Chen, Duan, & Mou, 2020). However, it is still an open problem for spectral CT reconstruction. Deep learning consists of two components: network architecture and objective function. The network architecture determines the complexity of the network model, while the objective function controls how to learn the model. Combination of these two components can improve the performances of deep learning methods. According to the use of label information in networks, deep learning based methods can be mainly divided into supervised, unsupervised and semi-supervised method. In this study, we focus on the supervised methods. Since the spectral CT images are correlated to x-ray energy bins and material components, it is difficult to prepare high quality labels datasets. Here, we first formulate high-quality of training dataset, which will lay a solid foundation for deep learning methods.

**B. Spectral CT Database**

Regarding deep learning based spectral CT reconstruction, how to formulate the database is very important. There are some typical factors that play an important role in spectral CT images, including the used x-ray spectrum, the energy bins thresholding, the composites of imaged objects. Here, we first elaborate on how to generate material component images. To ensure the proposed deep learning method can be employed in practice, instead of using phantoms, the material components images were obtained from clinical CT images rather than phantoms. Here, a set of DICOM images $\{\mathbf{X}_n\}_{n=1}^N$ (where $N$ represents the number of DICOM images) from "the 2016 Low-dose CT Grand Challenge" (AAPM) were collected as initial CT images. Some typical DICOM images were chosen as benchmark images. For a given benchmark image, the pixel values of $\mathbf{X}_n$ could be converted from CT values to linear attenuation coefficients image $\mathbf{Y}_n$. Assuming the set of $\{\mathbf{Y}_n\}_{n=1}^N$ only contains two materials for, i.e., bone and soft tissue. For generating the material images, we need to know the specified material components in each pixel. Here, all pixel values of $\mathbf{Y}_n$ can be segmented into bone $\mathbf{Y}_n^b$ and soft tissue $\mathbf{Y}_n^s$ components by using a soft threshold-based weighting method (Kyriakou, Meyer, Prell, & Kachelrieß, 2010). Again, the pixels below a certain threshold $T_s$ can be considered as soft tissue, while pixels above a higher threshold $T_b$ are treated as bone. Thus, the pixels between $T_s$ and $T_b$ are considered as the mixture of soft tissue and bone. Thus, a weighting function of bone component can be formulated as

$$\mathbf{w}^b(\mathbf{Y}_n(i,j)) = \begin{cases} 1, & \text{if } \mathbf{Y}_n(i,j) > T_b \\ 0, & \text{if } \mathbf{Y}_n(i,j) < T_s \\ \frac{\mathbf{X}_n(i,j) - T_s}{T_b - T_s}, & \text{otherwise} \end{cases}, \quad (4)$$

where $\mathbf{Y}_n(i,j)$ represents $(i,j)^{th}$ element of $\mathbf{Y}_n$. According to Eq. (4), $\mathbf{Y}_n^b$ and $\mathbf{Y}_n^s$ can be further written as

$$\mathbf{Y}_n^b = \mathbf{w}^b \mathbf{Y}_n, \quad (5)$$

and

$$\mathbf{Y}_n^s = (\mathbf{I} - \mathbf{w}^b)\mathbf{Y}_n, \quad (6)$$

where $\mathbf{I}$ represents the identity matrix.

To simulate spectral CT projection, we need to know the x-ray energy spectrum and the specified energy bins. Here, we can further assume that the x-ray energy spectrum is divided into $M$ energy bins. Regarding the spectral CT imaging for fan-beam geometry, the photon number along an x-ray path $\ell$ from the $m^{th}$ ($m=1,\ldots,M$) energy window $\Delta E_m = E_{m+1} - E_m$ can be read as the following expression

$$z_{m\ell} = \int_{E_m}^{E_{m+1}} I_{m\ell}(E) e^{\int_{\mathbf{r} \in \ell} -\mu(E,\mathbf{r})d\mathbf{r}} dE. \quad (7)$$

$\int_{E_m}^{E_{m+1}} dE$ and $\int_{r \in \ell} d\mathbf{r}$ represent the integral over the range of $\Delta E_m$ and along the x-ray path $\ell$. Since the imaged patients only contain two materials. $\mu(E, \mathbf{r})$ denotes the material linear attenuation coefficient for energy $E$ at position $\mathbf{r}$ and $I_{m\ell}(E)$ represents the original photon intensity emitting from the x-ray source for energy $E$. Since the imaging object only contains *two* materials (bone and soft tissue), $\mu(E, \mathbf{r})$ can be written as

$$\mu(E, \mathbf{r}) = \phi_b(E)\rho_b(\mathbf{r}) + \phi_s(E)\rho_s(\mathbf{r}), \tag{8}$$

$\phi_b(E)$ and $\phi_s(E)$ are the mass-attenuation coefficients of bone and soft tissue at energy $E$, which can be determined by searching the tables in the national institute of standards and technology (NIST) report (Hubbell & Seltzer, 1995). $\rho_1(\mathbf{r})$ and $\rho_2(\mathbf{r})$ represent the fraction of bone and soft tissue at location $\mathbf{r}$. We denote the original x-ray photon flux as $I_{m\ell}^{(0)}$ which can be expressed as

$$I_{m\ell}^{(0)} = \int_{E_m}^{E_{m+1}} I_{m\ell}(E) dE, \tag{9}$$

Substituting Eqs. (8)-(9) into Eq. (7), we can obtain

$$e_{m\ell} = \int_{E_m}^{E_{m+1}} r_{m\ell}(E) e^{-\int_{\mathbf{r} \in \ell} \phi_b(E)\rho_b(\mathbf{r}) + \phi_s(E)\rho_s(\mathbf{r}) d\mathbf{r}} dE, \tag{10}$$

$r_{m\ell}(E) = I_{m\ell}(E)/I_{m\ell}^{(0)}$ represents the normalized energy spectrum distribution of x-ray intensity and detector sensitivity and $e_{m\ell} = z_{m\ell}/I_{m\ell}^{(0)}$. Finally, we can obtain the spectral CT projections by performing a logarithm operation on both sides of Eq. (10) and then we have

$$p_{m\ell} = \ln \int_{E_m}^{E_{m+1}} r_{m\ell}(E) e^{-\int_{\mathbf{r} \in \ell} \phi_b(E)\rho_b(\mathbf{r}) + \phi_s(E)\rho_s(\mathbf{r}) d\mathbf{r}} dE, m = 1, \dots, M \text{ and } \ell = 1, \dots, L \tag{11}$$

Now, given a number of CT images, we can generate the spectral CT projections and then reconstruct images by using the FBP method.

Figure 1 summarizes the methods of the generation of the spectral CT database. We can see the generation flowchart can mainly divided into three stages: material segmentation, projection and reconstruction. We can separate the conventional CT images into specified material maps, such as bone and soft tissue. Then, we can generate the spectral CT noise-free and noisy projections by depending on x-ray emitting spectral and material-specified coefficients. Finally, we can reconstruct the spectral CT images using reconstruction methods, such as FBP.

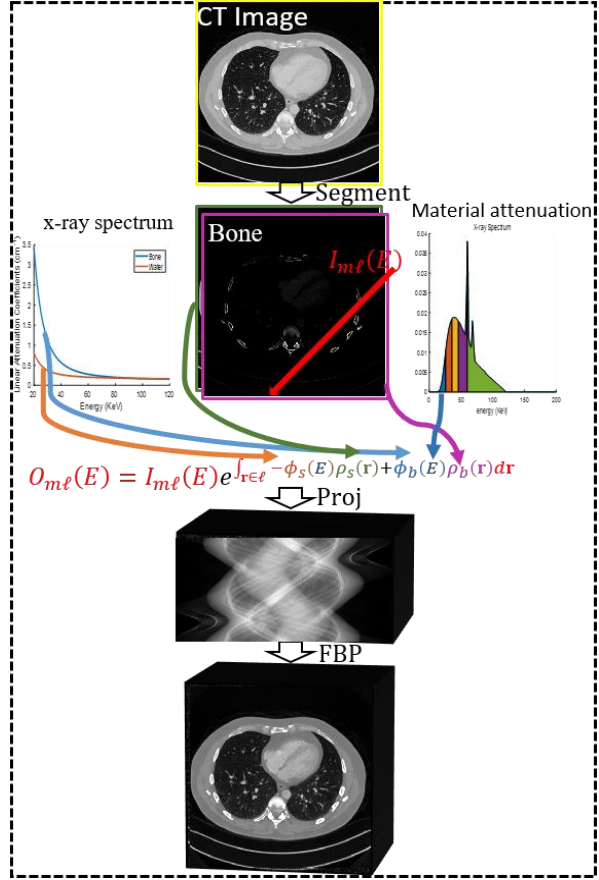

Figure 1 The generation flowchart of spectral CT database

### III. ULTRA Network

#### A. Motivation

U-net is a typical CNN network, which is first proposed for image segmentation (Ronneberger, Fischer, & Brox, 2015). Then, it was rapidly applied to image blurring (Tao, Gao, Shen, Wang, & Jia, 2018), image enhancement (L. Zhang, Ji, Lin, & Liu, 2017), image recovery (Ma, Shu, Bai, Wang, & Samaras, 2018), artifact reduction (C. Zhang & Xing, 2018), image denoising (Heinrich, Stille, & Buzug, 2018). To further improve the performance of the original U-net on CT image denoising, a skip connection was added between input and output and then generate FBPConvNet (Jin, McCann, Froustey, & Unser, 2017). Although FBPConvNet has obtained great success in CT image denoising, there remain areas for improvement. First, the FBPCovNet can encode image feature by using multichannel filters, such that there are multiple feature maps at each layer. Although the multi-scale convolution layers can reduce the noise within a noisy image, the features and details are also missing. Besides, image features are only extracted by top-to-bottom net flow, which can result in the image features from different scale not being shared fully. It further causes image features to be missing during the process of feature encoding. Second, FBPCovNet can operate global normalization on entire database, which can relax the independency of each batch. Again, the unique features of image-self and noise distribution may not be fully explored in FBPCovNet. Third, the deep learning CT denoising methods mainly focus on $L_2^2$ loss. Because $L_2^2$ loss can cause image edge blurred by over-smoothing, which can compromise reconstructed image quality. Fourth, the prior knowledge of medical images has been used extensively in the fields of image processing and computer vision, most of deep learning-based CT reconstruction method ignores this issue. To overcome these issue, we first propose a new network architecture and then formulate a new loss function based on $L_p^p$ $(p > 0)$ and prior knowledge.

**B. Network Architecture**

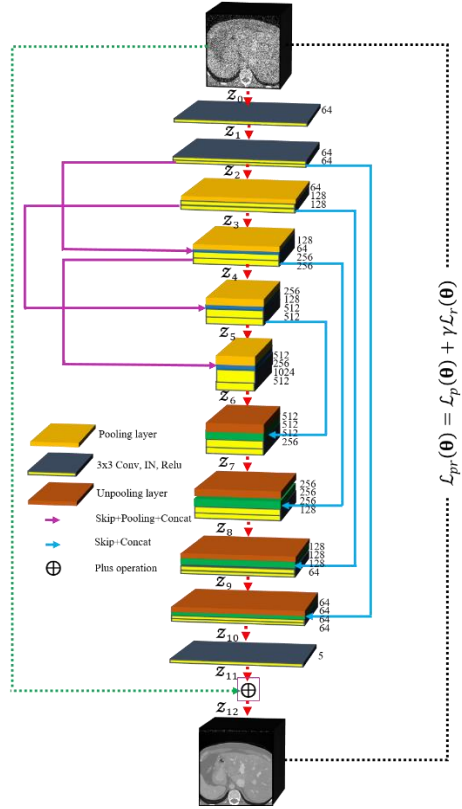

Figure 2 The proposed RESULT network architecture

Compared with the previous efforts, the features of our ULTRA network can be mainly embodied from the following three points:

*1) Various Multi-scale Feature Fusion:* Both U-net and other modified U-net version (including FBPCovNet) are concentrating on feature extraction by convolution layers with different filter sizes. This means the convolution features from previous convolution layers may lose in later convolution lay. Again, some of the history convolution features will be missing so that U-net may not be fully encoded in the input images features. To overcome these issues, we further introduce the output of the early convolution layer into the later convolution layer via skip connection. In this way, it can enhance the encoding ability of feature for the network. Specifically, the input size of spectral CT images is $\mathcal{Z}_0 \in \mathcal{R}^{128 \times 128 \times 5}$, and it can be coupled into the proposed network with $\mathcal{Z}_1 \in \mathcal{R}^{128 \times 128 \times 64}$. In this stage, the features of spectral CT images from different energy bins are fused, this means the correlation between different energy bins can be explored collaboratively. $\mathcal{Z}_1$ is followed by two convolution layers and we can obtain $\mathcal{Z}_2$. The key point is $\mathcal{Z}_2$ can be divided into two paths: one is the input of $\mathcal{Z}_3$ and the other is the input of $\mathcal{Z}_4$ after pooling layer. $\mathcal{Z}_3$ and $\mathcal{Z}_4$ are very similar to $\mathcal{Z}_2$. Such a strategy is the benefit of re-using multi-scale image feature.

*2) Multichannel Filtering Enhancement:* Increasing the number of channels is a standard approach to enhance the expressive power of the network. The multiple channels also have an analog in iterative methods: Compared with conventional U-net based networks, there are more channels are treated as input of $\mathcal{Z}_4 - \mathcal{Z}_6$. The architecture further generalizes this by introducing more filters to make combinations of various channels.

*3) Sparsity Constraint:* The biggest challenge for U-net based network (for example, FBPConvNet) is that it can easily cause oversmoothing. The common strategies avoid overfitting by using data augmentation, simpler model structure and introducing regularization. In this study, we introduce $L_p^p$ ($p>0$) to model the loss function, which can reduce the overfitting by selecting an appropriate $p$ value. Besides, residual learning can also benefit to reduce overfitting to some extent.

*4) Prior Knowledge regularization:* The iterative spectral CT reconstruction methods based on prior knowledge are becoming popular and successful. These methods provide clear image edges and other details by incorporating image characteristics of spectral CT into the reconstruction model. However, such beneficial knowledge is missing in many current deep reconstruction networks, including FBPConvNet. In this network, we have incorporated such established prior into our ULTRA network to improve the stability while maximizing image quality. Specifically, we have adopted the anisotropically weighted total variation to characterize the sparsity in the spatial-spectral domain to enhance the loss function. It can benefit to open the window of prior knowledge based deep learning for spectral CT reconstruction.

## C. Prior Knowledge Driven $L_p^p(p>0)$ Loss

Conventional U-net based CT imaging methods mainly focus on minimizing mean square error (MSE, i.e., $L_2^2$ loss) between input and output. Although the $L_2^2$ loss can achieve success in CT imaging, it can still cause image edge blurring and missing finer detail structures. In this study, we propose a general $L_p^p(p>0)$ loss function.

Given a spectral CT image pair $\{\mathcal{O}, \mathcal{Q}\}$, where $\mathcal{O}$ is the noisy input image and $\mathcal{Q}$ corresponds label, the $L_p^p$ loss of each pixel can be denoted as

$$\mathcal{L}_p(\boldsymbol{\theta}) = \frac{1}{N_h \times N_w \times M} \sum_{w=1}^{N_w} \sum_{h=1}^{N_h} \sum_{m=1}^{M} \|\Phi(\mathcal{O}_{whm}, \boldsymbol{\theta}) - \mathcal{Q}_{whm}\|_p^p, \quad (12)$$

where $\Phi$ is the learned network (parameters) for estimating the input and $\boldsymbol{\theta}$ represents the learned parameters of the network. $\mathcal{O}_{whm}$ and $\mathcal{Q}_{whm}$ represent $(w, h, m)^{th}$ pixel value of $\mathcal{O}$ and $\mathcal{Q}$. $p>0$ represent different loss. Especially, $p=1$ and $p=2$ represent the $L_1$ and $L_2^2$ losses. In this study, we mainly focus on $0<p<2$. When the number of free parameters within the network $\Phi$ is large and the amount of labeled data is small in comparison, the overfitting will appear and further result in poor performance. Indeed, one feasible strategy is to increase the number of labels, although this will increase human efforts. Besides, since the x-ray emitting spectrum is difficult to model, which result in the inability to obtain ground labels in practice. The traditional approach is by penalizing the $L_2$ norm of the weight vector. However, this kind of regularization constraints are very general and cannot exploit the characteristics of the spatial structures implying reconstructed spectral CT images.

Considering the spatial structure of output, the loss function based on prior knowledge can be formulated as

$$\mathcal{L}_r(\boldsymbol{\theta}) = \frac{1}{N_h \times N_w \times M} \sum_{w=1}^{N_w} \sum_{h=1}^{N_h} \sum_{m=1}^{M} \mathcal{G}(\psi(\mathcal{O}_{whm})), \quad (13)$$

where $\psi(\mathcal{O}_{whm}) \in \mathcal{R}^{n_1^3}$ is a vector containing $N_2 \times N_2 \times N_2$ neighborhood pixels of $\Phi(\mathcal{O}_{whm})$. $\mathcal{G}$ is a function converting a vector of the size $N_2^3$ to one pixel. This means $\mathcal{G}$ can characterize the spatial property of the expected output. In fact, Eq. (13) can be considered as an unsupervised loss. Especially, if the labels are treated as unlabeled data, only the unsupervised loss is considered for training the network.

To clarify this point, the $\psi(\mathcal{O}_{whm})$ can be written as

$$\psi(\mathcal{O}_{whm}) = \begin{bmatrix} \Phi_1(\mathcal{O}_{whm}) \\ \Phi_2(\mathcal{O}_{whm}) \\ \vdots \\ \Phi_{N_2^3}(\mathcal{O}_{whm}) \end{bmatrix}, \quad (14)$$

where $\Phi_i(\mathcal{O}_{whm}) = \Phi(\mathcal{O}_{whm+i})$ and $(n_2 = 1, \dots, N_2^3)$. In this way, $\mathcal{L}_r(\Phi)$ can be written as a function of

$$\mathcal{L}_r(\boldsymbol{\theta}) = \frac{1}{N_h \times N_w \times M} \sum_{w=1}^{N_w} \sum_{h=1}^{N_h} \sum_{m=1}^{M} \mathcal{G}\left(\Phi_1(\mathcal{O}_{whm}), \dots, \Phi_{N_2^3}(\mathcal{O}_{whm})\right). \quad (15)$$

Regarding the minimization of Eq. (15), we only discuss the objection function is differentiate. Here, the gradient descent based on multivariate chain rule is employed and then we have

$$\frac{\partial \mathcal{L}_r(\boldsymbol{\theta})}{\partial \boldsymbol{\theta}} = \frac{1}{N_h \times N_w \times M} \sum_{w=1}^{N_w} \sum_{h=1}^{N_h} \sum_{m=1}^{M} \sum_{n_2=1}^{N_2^3} \frac{\partial \mathcal{L}_r(\boldsymbol{\theta})}{\partial \Phi_{n_2}(\mathcal{O}_{whm})} \times \frac{\partial \Phi_{n_2}(\mathcal{O}_{whm})}{\partial \boldsymbol{\theta}}, \quad (16)$$

where $\frac{\partial \Phi_{n_2}(\mathcal{O}_{whm})}{\partial \boldsymbol{\theta}}$ represents the derivative of $\Phi(\mathcal{O}_{whm})$ about $\boldsymbol{\theta}$ at the location of $\mathcal{O}_{whm+n_2}$. $\frac{\partial \mathcal{L}_r(\boldsymbol{\theta})}{\partial \Phi_{n_2}(\mathcal{O}_{whm})}$ is the coefficient from the specified loss function. Again, the backpropagation of the loss function is combined by the derivatives of $\Phi$ with respect to $\boldsymbol{\theta}$. The coefficients of this linear combination are dependent on the definition of the loss function. The proposed $\mathcal{L}_r(\boldsymbol{\theta})$ can be considered as a combination of $L_p$ loss and prior knowledge and its general format can be given as

$$\mathcal{L}_{pr}(\boldsymbol{\theta}) = \mathcal{L}_p(\boldsymbol{\theta}) + \gamma \mathcal{L}_r(\boldsymbol{\theta}), \quad (17)$$

where $\gamma > 0$ represents the proposition of two terms.

The piecewise smoothness property of spectral CT images has been explored extensively. Total variation is usually employed to characterize this property. It is first employed in CT reconstruction (Sidky & Pan, 2008) and then was developed for interior spectral CT reconstruction(Xu, et al., 2012). It can remove noise with most of of the image detail preserved. Since the channel-wise spectral CT images are coming from the same imaged object, these images are correlated. Thus, we propose a $\mathcal{L}_r(\boldsymbol{\theta})$ based on anisotropy weight total variation to characterize the spatial information of network output and it can be given as

$$\mathcal{L}_r(\boldsymbol{\theta}) = \frac{1}{N_h \times N_w \times M} \sum_{w=1}^{N_w} \sum_{h=1}^{N_h} \sum_{m=1}^{M} \left( \beta_1 |\Phi(\mathcal{O}_{whm}, \boldsymbol{\theta}) - \Phi(\mathcal{O}_{(w-1)hm}, \boldsymbol{\theta})| + \beta_2 |\Phi(\mathcal{O}_{whm}, \boldsymbol{\theta}) - \Phi(\mathcal{O}_{w(h-1)m}, \boldsymbol{\theta})| + \beta_3 |\Phi(\mathcal{O}_{whm}, \boldsymbol{\theta}) - \Phi(\mathcal{O}_{wh(m-1)}, \boldsymbol{\theta})| \right), \quad (18)$$

Substituting Eqs. (12) and (18) into Eq. (17), the total loss can be expressed as

$$\mathcal{L}_{pr}(\boldsymbol{\theta}) = \frac{1}{N_h \times N_w \times M} \sum_{w=1}^{N_w} \sum_{h=1}^{N_h} \sum_{m=1}^{M} \{ \|\Phi(\mathcal{O}_{whm}, \boldsymbol{\theta}) - \mathcal{Q}_{whm}\|_p^p + \gamma \beta_1 |\Phi(\mathcal{O}_{whm}, \boldsymbol{\theta}) - \Phi(\mathcal{O}_{(w-1)hm}, \boldsymbol{\theta})| + \gamma \beta_2 |\Phi(\mathcal{O}_{whm}, \boldsymbol{\theta}) - \Phi(\mathcal{O}_{w(h-1)m}, \boldsymbol{\theta})| + \gamma \beta_3 |\Phi(\mathcal{O}_{whm}, \boldsymbol{\theta}) - \Phi(\mathcal{O}_{wh(m-1)}, \boldsymbol{\theta})| \}, \quad (19)$$

By incorporating the definition of $L_p^p$ loss, Eq. (19) can be read as

$$\mathcal{L}_{pr}(\boldsymbol{\theta}) = \frac{1}{N_h \times N_w \times M} \sum_{w=1}^{N_w} \sum_{h=1}^{N_h} \sum_{m=1}^{M} \{ |\Phi(\mathcal{O}_{whm}, \boldsymbol{\theta}) - \mathcal{Q}_{whm}|^p + \gamma \beta_1 |\Phi(\mathcal{O}_{whm}, \boldsymbol{\theta}) - \Phi(\mathcal{O}_{(w-1)hm}, \boldsymbol{\theta})| + \gamma \beta_2 |\Phi(\mathcal{O}_{whm}, \boldsymbol{\theta}) - \Phi(\mathcal{O}_{w(h-1)m}, \boldsymbol{\theta})| + \gamma \beta_3 |\Phi(\mathcal{O}_{whm}, \boldsymbol{\theta}) - \Phi(\mathcal{O}_{wh(m-1)}, \boldsymbol{\theta})| \}, \quad (20)$$

Here, we only discuss $0 < p < 2$, it is not to differentiate. The subgradient method can generalize the derivative to non-differentiable functions by computing

$$\frac{\partial \mathcal{L}_{pr}(\boldsymbol{\theta})}{\partial \boldsymbol{\theta}} = \frac{1}{N_h \times N_w \times M} \sum_{w=1}^{N_w} \sum_{h=1}^{N_h} \sum_{m=1}^{M} \Bigg\{ psign(\Phi(\mathcal{O}_{whm}, \boldsymbol{\theta}) - \mathcal{Q}_{whm}) |\Phi(\mathcal{O}_{whm}, \boldsymbol{\theta})$$

$$- \mathcal{Q}_{whm}|^{p-1} \frac{d(\Phi(\mathcal{O}_{whm}, \boldsymbol{\theta}) - \mathcal{Q}_{whm})}{d\boldsymbol{\theta}}$$

$$+ \gamma \Bigg( \beta_1 sign \left( \Phi(\mathcal{O}_{whm}, \boldsymbol{\theta}) - \Phi(\mathcal{O}_{(w-1)hm}, \boldsymbol{\theta}) \right) \frac{d \left( \Phi(\mathcal{O}_{whm}, \boldsymbol{\theta}) - \Phi(\mathcal{O}_{(w-1)hm}, \boldsymbol{\theta}) \right)}{d\boldsymbol{\theta}}$$

$$+ \beta_2 sign \left( \Phi(\mathcal{O}_{whm}, \boldsymbol{\theta}) - \Phi(\mathcal{O}_{w(h-1)m}, \boldsymbol{\theta}) \right) \frac{d \left( \Phi(\mathcal{O}_{whm}, \boldsymbol{\theta}) - \Phi(\mathcal{O}_{w(h-1)m}, \boldsymbol{\theta}) \right)}{d\boldsymbol{\theta}}$$

$$+ \beta_3 sign \left( \Phi(\mathcal{O}_{whm}, \boldsymbol{\theta}) - \Phi(\mathcal{O}_{wh(m-1)}, \boldsymbol{\theta}) \right) \frac{d \left( \Phi(\mathcal{O}_{whm}, \boldsymbol{\theta}) - \Phi(\mathcal{O}_{wh(m-1)}, \boldsymbol{\theta}) \right)}{d\boldsymbol{\theta}} \Bigg) \Bigg\}$$

,(21)

where $sign(\cdot)$ represents the sign function and $sign(\Phi(\mathcal{O}_{whm}, \boldsymbol{\theta}) - \mathcal{Q}_{whm}) = \frac{\Phi(\mathcal{O}_{whm}, \boldsymbol{\theta}) - \mathcal{Q}_{whm}}{|\Phi(\mathcal{O}_{whm}, \boldsymbol{\theta}) - \mathcal{Q}_{whm}|}$.

According to Eq. (21), when $1 < p < 2$, it is reasonable. However, $0 < p < 1$, the $|\Phi(\mathcal{O}_{whm}, \boldsymbol{\theta}) - \mathcal{Q}_{whm}|$ may be equal to zero and then result in the singularity of Eq. (21). To address this case, Eq. (21) can be rewritten as

$$\frac{\partial \mathcal{L}_{pr}(\boldsymbol{\theta})}{\partial \boldsymbol{\theta}} = \frac{1}{N_h \times N_w \times M} \sum_{w=1}^{N_w} \sum_{h=1}^{N_h} \sum_{m=1}^{M} \Bigg\{ psign(\Phi(\mathcal{O}_{whm}, \boldsymbol{\theta}) - \mathcal{Q}_{whm}) (|\Phi(\mathcal{O}_{whm}, \boldsymbol{\theta}) - \mathcal{Q}_{whm}|$$

$$+ \varepsilon)^{p-1} \frac{d(\Phi(\mathcal{O}_{whm}, \boldsymbol{\theta}) - \mathcal{Q}_{whm})}{d\boldsymbol{\theta}}$$

$$+ \gamma \Bigg( \beta_1 sign \left( \Phi(\mathcal{O}_{whm}, \boldsymbol{\theta}) - \Phi(\mathcal{O}_{(w-1)hm}, \boldsymbol{\theta}) \right) \frac{d \left( \Phi(\mathcal{O}_{whm}, \boldsymbol{\theta}) - \Phi(\mathcal{O}_{(w-1)hm}, \boldsymbol{\theta}) \right)}{d\boldsymbol{\theta}}$$

$$+ \beta_2 sign \left( \Phi(\mathcal{O}_{whm}, \boldsymbol{\theta}) - \Phi(\mathcal{O}_{w(h-1)m}, \boldsymbol{\theta}) \right) \frac{d \left( \Phi(\mathcal{O}_{whm}, \boldsymbol{\theta}) - \Phi(\mathcal{O}_{w(h-1)m}, \boldsymbol{\theta}) \right)}{d\boldsymbol{\theta}}$$

$$+ \beta_3 sign \left( \Phi(\mathcal{O}_{whm}, \boldsymbol{\theta}) - \Phi(\mathcal{O}_{wh(m-1)}, \boldsymbol{\theta}) \right) \frac{d \left( \Phi(\mathcal{O}_{whm}, \boldsymbol{\theta}) - \Phi(\mathcal{O}_{wh(m-1)}, \boldsymbol{\theta}) \right)}{d\boldsymbol{\theta}} \Bigg) \Bigg\}$$

,(22)

where $\varepsilon$ is a small positive value. In this study, it is setting as $10^{-4}$. The gradient of the anisotropy weight total variation penalty at one location is defined by a linear combination of gradient values from three connected neighborhood pixels of different direction. This implies dependency between pixels in the output image, the benefit of which is the goal of weight total variation penalty. In fact, the image gradient from the horizontal and vertical directions are equal, i.e., $\beta_1 = \beta_2$. In this case, Eq. (22) can be involved into

$$\frac{\partial \mathcal{L}_{pr}(\boldsymbol{\theta})}{\partial \boldsymbol{\theta}} = \frac{1}{N_h \times N_w \times M} \sum_{w=1}^{N_w} \sum_{h=1}^{N_h} \sum_{m=1}^{M} \Bigg\{ p\, sign(\Phi(\mathcal{O}_{whm}, \boldsymbol{\theta}) - \mathcal{Q}_{whm})(|\Phi(\mathcal{O}_{whm}, \boldsymbol{\theta}) - \mathcal{Q}_{whm}|$$
$$+ \varepsilon)^{p-1} \frac{d(\Phi(\mathcal{O}_{whm}, \boldsymbol{\theta}) - \mathcal{Q}_{whm})}{d\boldsymbol{\theta}}$$
$$+ \gamma \Bigg( sign\left(\Phi(\mathcal{O}_{whm}, \boldsymbol{\theta}) - \Phi(\mathcal{O}_{(w-1)hm}, \boldsymbol{\theta})\right) \frac{d\left(\Phi(\mathcal{O}_{whm}, \boldsymbol{\theta}) - \Phi(\mathcal{O}_{(w-1)hm}, \boldsymbol{\theta})\right)}{d\boldsymbol{\theta}}$$
$$+ sign\left(\Phi(\mathcal{O}_{whm}, \boldsymbol{\theta}) - \Phi(\mathcal{O}_{w(h-1)m}, \boldsymbol{\theta})\right) \frac{d\left(\Phi(\mathcal{O}_{whm}, \boldsymbol{\theta}) - \Phi(\mathcal{O}_{w(h-1)m}, \boldsymbol{\theta})\right)}{d\boldsymbol{\theta}}$$
$$+ w\, sign\left(\Phi(\mathcal{O}_{whm}, \boldsymbol{\theta}) - \Phi(\mathcal{O}_{wh(m-1)}, \boldsymbol{\theta})\right) \frac{d\left(\Phi(\mathcal{O}_{whm}, \boldsymbol{\theta}) - \Phi(\mathcal{O}_{wh(m-1)}, \boldsymbol{\theta})\right)}{d\boldsymbol{\theta}} \Bigg) \Bigg\}$$
,(23)

where $w = \beta_3/\beta_2$. According to Eq. (23), we can change the proposition of variation from three directions by adjusting the weight $w$ in practice.

**IV. Experimental Results**

In this section, we first introduce the spectral CT datasets for training and evaluating the proposed networks, including data preparation and hyperparameters selection. Then, the numerical, preclinical mouse and physical phantom experiments are employed to evaluate the performance of our proposed network by comparing with total variation minimization (TVM) (Xu, et al., 2012), other classic deep learning based inverse imaging network, i.e. FBPConvNet (Jin, et al., 2017). The experiment results demonstrate the advantages of our proposed network for spectral CT imaging. The following experimental results will be divided into numerical experiments, preclinical mice and physical phantom.

**A. Experimental Setup**
*A.1 Datasets*
The 2016 NIH-AAPM-Mayo Clinic Low Dose CT Grand Challenge contains 10 anonymized patient normal dose abdominal CT images (acquired at 120 kV and 200 effective mAs), which are used to create a database for traditional CT. Regarding the scan configuration, the distance between x-ray source and detector is set as 560mm and the length of PCD unit can cover a length of 0.225mm. The projection views are set as 1080. The functions of *fanbeam* and *ifanbeam* are used to generate projections and reconstruct spectral CT images. In this study, a 120KVp x-ray source is employed, where the x-ray spectrum can be divided into five energy bins, as shown in **Fig. 3(a)**. To balance the photon number of each energy bin is almost equal, the specified energy bins are [19 26], [26 36], [36 46], [46 60] and [60 120]. Since the photon counting detector from MARS setup is 12-bit, the number of x-ray path is set as 4096. The Poisson noise is applied into the photon-domain projections to obtain noisy projections.

Here, the training images are mainly extracted from 8 patients chosen as our original CT images in AAPM to generate the spectral CT projections by following the flowchart of **Fig. 1**. Here, the number of original CT images is 1100. Then, the FBP is employed to obtain noise-free and noisy reconstruction results, where noisy and noise-free are treated as input and output of networks. 200 original CT images from the other 2 patients are employed to generate test datasets by using the same procedure. For the network training, data augmentation technique is utilized, which can improve the performance of networks. Specifically, a total of 43400 volumes with the size of 128x128x5 are used to train all networks. For the numerical experiments, root mean square error (RMSE) and structure similarity (SSIM) are employed to quantitative assess the performance of all algorithms.

*A.2 Network Training*

Regarding network training, Adam method is employed to optimize all of the networks (Kingma & Ba, 2014), where with $\alpha_1 = 0.9$ and $\alpha_2 = 0.999$ in the loss function of our proposed network. To avoid the inconsistency of size between all feature maps and the input, we padded zeros around the boundaries before the convolution. The learning rate is decreased along the number of epochs. Here, the number of epoch is set as 50 and the learning rate is shown in **Fig. 3(b)**. The batch size is set as 32. For the value of $p$ and regularization parameter $\gamma$ are set as 1.2 and 0.001 in $\mathcal{L}_{pr}(\boldsymbol{\theta})$. Regarding as the parameter of w, it is set as 0.1.

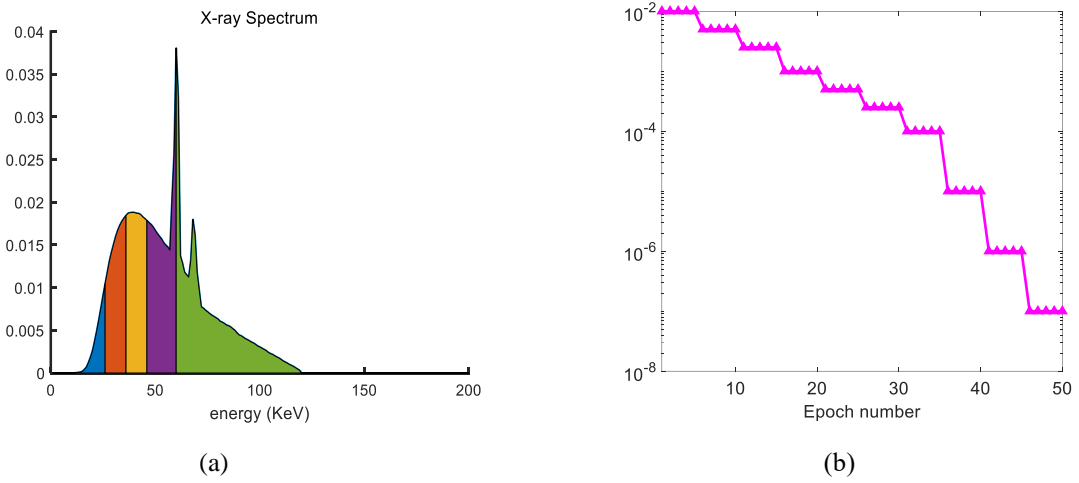

Fig. 3 (a) represent x-ray energy spectrum and (b) represents the learn rate vs epoch number

**B. Experimental Results**

*B.1 Numerical Simulation*

**Figures 4-6** shows reconstructed spectral CT images from three cases with five energy bins using all algorithms. It can be seen that FBP results most amount of noise so that most image details and edges are obscured. Compared with the FBP results, TVM can improve image quality with some image features and details by incorporating the sparsity prior into iteration reconstruction models. In this study, TVM is stopped after 50 iterations. However, this iteration method needs more time to obtain the final reconstruction results. It is feasible to implement fan-beam geometry in practice. Although the power of computational is improved nowadays, there is still a big challenge for performing such regularization based iteration methods for helical spectral CT. Unfortunately, the basic iteration methods is still employed in commercial medical CT systems, such as the advanced MARS spectral CT scanner. Besides, the image edges and profiles are blurred so that it is difficult to discriminate clear structure. Compared with traditional methods, the FBPConvNet can improve image quality a lot in terms of image details recovery and features preservation. However, some fine image details are missing or obscured. Compared with FBPConvNet method, our proposed network can recover spectral CT images with much more image details and edges with an attractive capability in capturing smaller image structures and details.

To further demonstrate the advantages of the proposed network, two regions of interests (ROIs) were extracted and magnified version are also given in **Fig. 4**. Regarding as the image structure indicated by arrow "**1**", it was not as well visualised in other comparisons of all energy bins, including the FBPConvNet. However, this structure was well preserved in our proposed ULTRA results. Compared with the FBPConvNet, the proposed method could recover better results, which could be confirmed by the region marked with arrow "**2**". According to the indicated image edges by "**3**" and "**4**", we could observe that the proposed ULTRA results could provide clearer image edges than other methods. Besides, other two ROIs were extracted in **Fig. 5**. It could be seen that image edge marked with arrow "**5**" from TVM was deformed by suppressing noise. Although it could berecovered by FBPConvNet to some extent, the image edges were still blurred. Compared with TVM and FBPConvNet, our method could achieve well delineated image edge. It could be seen from **Fig. 5** that the details with arrow "**6**" cannot be discriminated from the results of FBP, TVM and FBPConvNet. Fortunately, it could be clearly observed from our proposed ULTRA results. The similar conclusions could also be reached from the image structures by "**7**" and "**8**". Finally, another two ROIs were also extracted from case 3# and magnified in **Fig. 6**. From the results of ROIs, we observed that the proposed network obtained higher quality reconstructed images with much image features and details with artifacts reduction than other comparisons, as indicated by image structures with arrows "**9-12**". All these cases and extracted ROIs further validate the outperformance of our proposed ULTRA network.

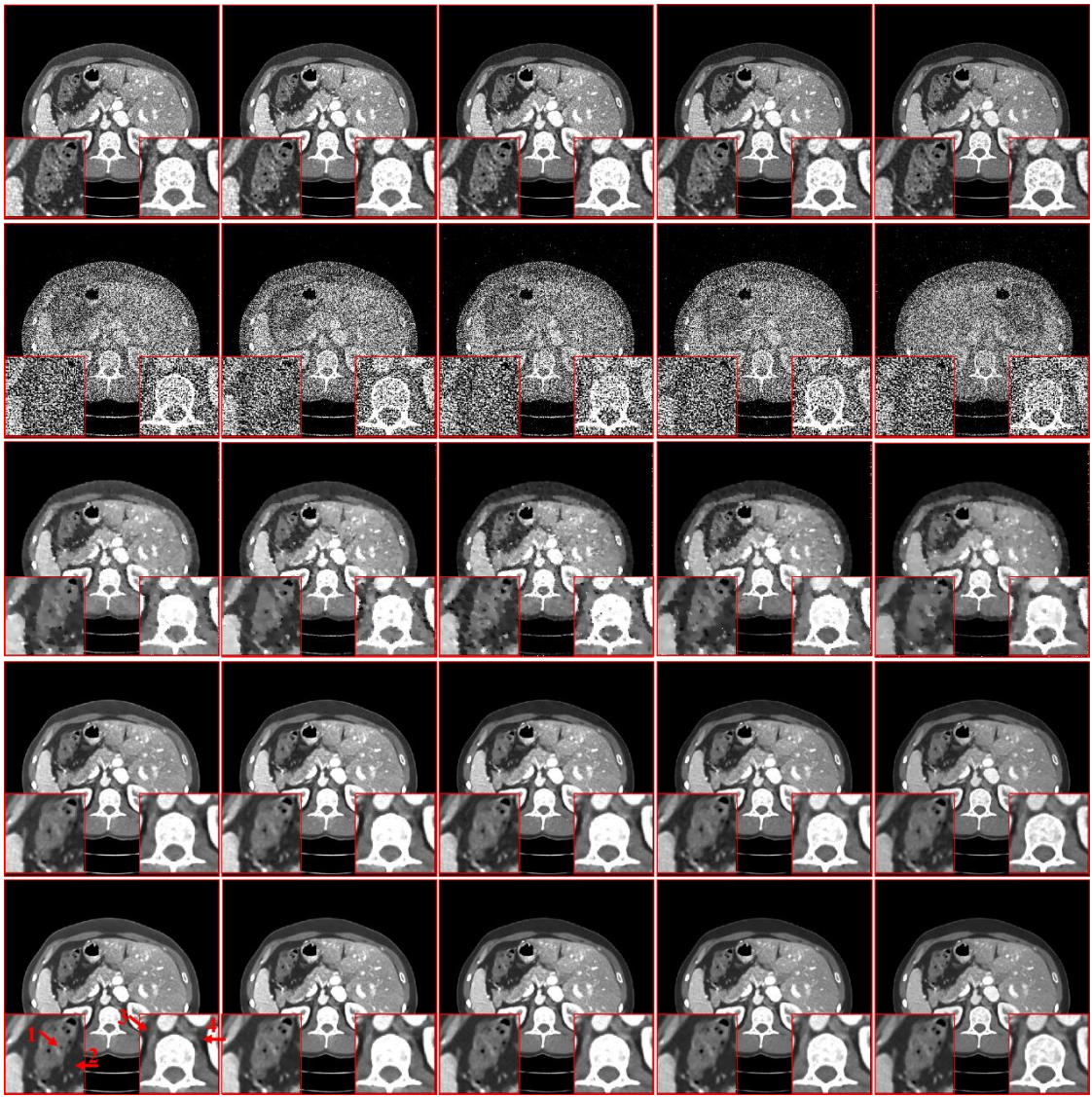

Fig. 4 The reconstruction results of case 1. The 1st-5th rows represent ground truth, FBP results with noisy projection, TVM, FBPConvNet and the proposed network. The 1st-5th columns represent reconstructed results from 1st-5th energy bins. The display window is [-200 400] HU.

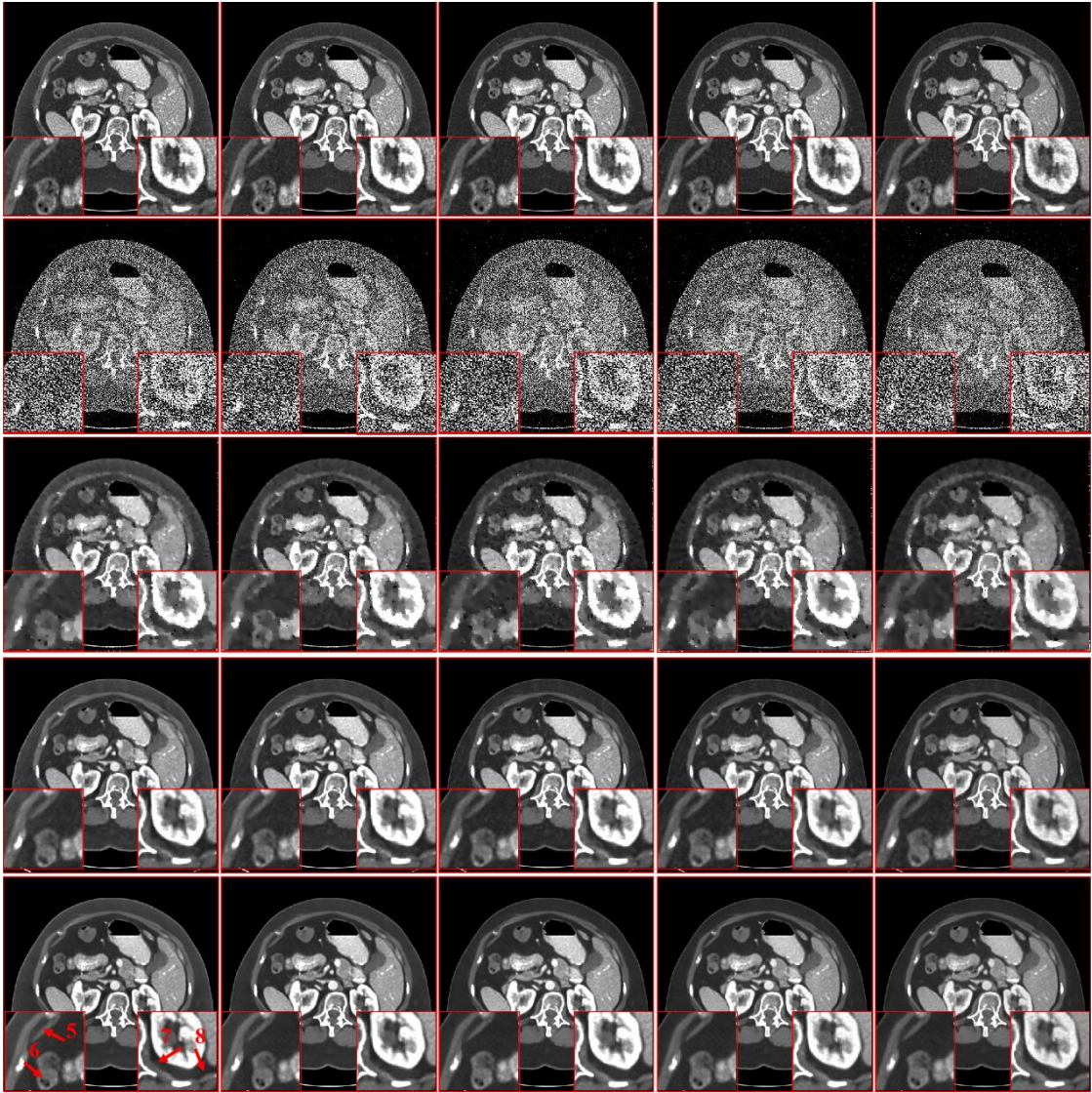

Fig. 5 Same as Fig. 4 but from case 2.

To make quantitative comparisons for all methods, the quantitative results of the reconstructed images of three representative cases from all energy bins are listed in **Table I**. Overall, we can observe that the deep learning-based network methods can obtain better results than traditional methods, including FBP and TVM methods. In addition, **Table I** demonstrates our proposed ULTRA network can obtain better results than classical FBPConvNet. Specifically, the proposed ULTRA network can obtain the smallest RMSEs for all energy bins than FBPConvNet, followed by TVM and FBP. The SSIM can measure the similarity between the reconstructed images and references, which are recently employed to compare reconstructed CT image quality. In terms of SSIM, the closer to 1.0 the value is, the better the reconstructed image quality is. According to the Table **I**, it can be seen that the proposed ULTRA network can achieve the greatest SSIM values for all energy bins than other reconstruction methods. The quantitative results validate the advantages of the proposed ULTRA network with providing better results.

In this study, all the source code of deep learning network were programmed with Python and implemented using the TensorFlow library on a NVIDA TITAN XP GPU. All programs were implemented on a PC (16 CPUs @3.70GHz, 32.0GB RAM and 8.0GB VRAM) with Windows 10. Regarding the computational cost, each iterations for TVM, FBPConvNet and ULTRA methods consumed 9.27, 2312.61, 9.53 and 9.55 seconds. Obviously, the iteration type reconstruction method needed more time than deep learning-based methods. In fact, the computation costs of deep learning-based reconstruction methods were comparable with traditional analytic methods.

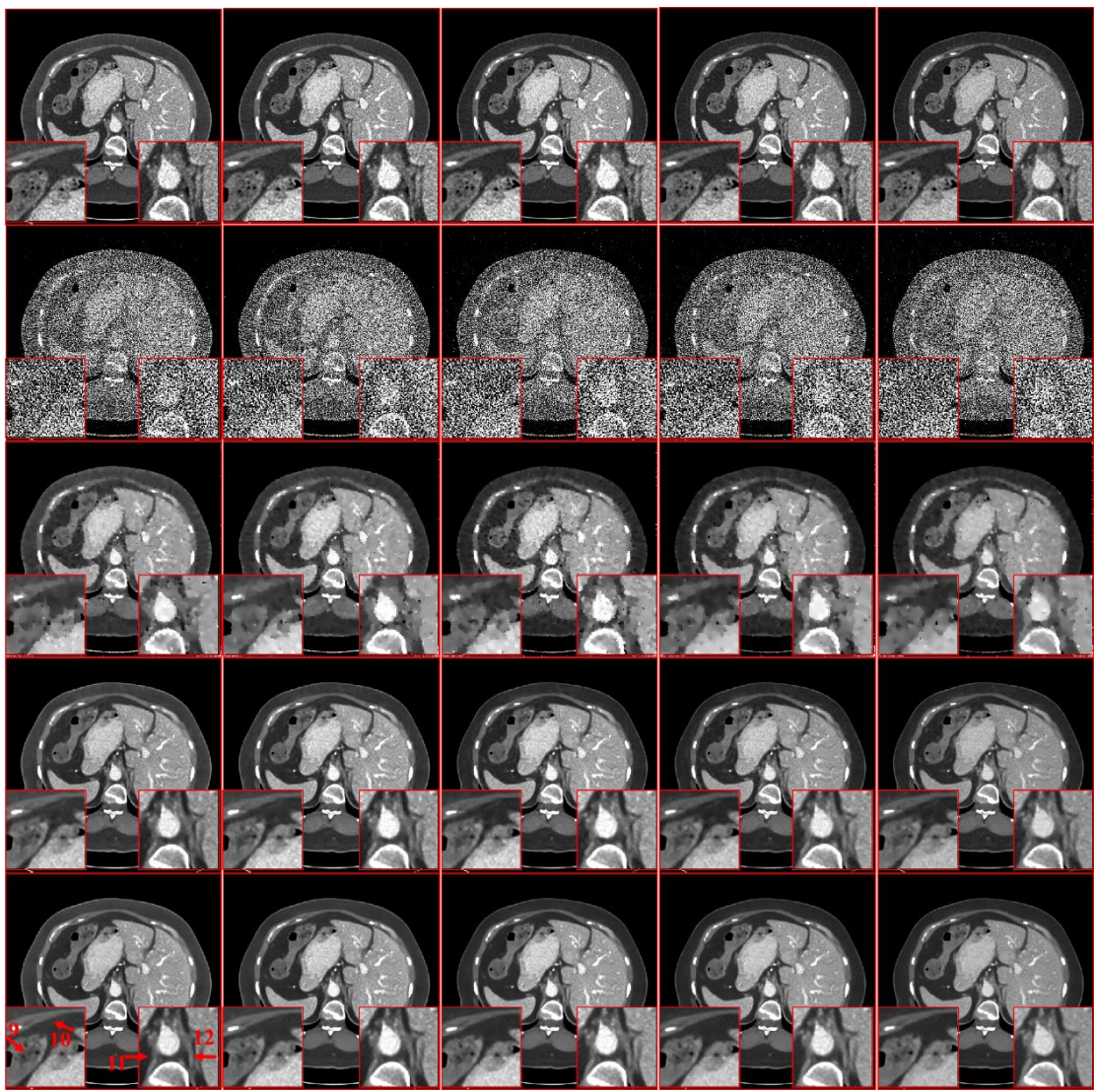

Fig. 6 Same as Fig. 4 but from case 3.

Table I. Quantitative evaluation results of the reconstructed images of representative three cases

| Cases | Metrics | Methods | 1st channel | 2nd channel | 3rd channel | 4th channel | 5th channel |
|---|---|---|---|---|---|---|---|
| #1 | RMSE | FBP | 329.5626 | 328.9475 | 380.9333 | 383.5904 | 397.7566 |
| | | TVM | 83.3381 | 85.1668 | 91.0334 | 87.4181 | 84.3508 |
| | | FBPConvNet | 33.7083 | 35.3752 | 34.4239 | 31.5113 | 28.4086 |
| | | ULTRA | **33.1894** | **34.7159** | **33.8983** | **31.0086** | **27.5965** |
| | SSIM | FBP | 0.3764 | 0.3778 | 0.3323 | 0.3185 | 0.2954 |
| | | TVM | 0.8774 | 0.8792 | 0.8619 | 0.8623 | 0.8624 |
| | | FBPConvNet | 0.9390 | 0.9380 | 0.9384 | 0.9407 | 0.9415 |

|     |      | ULTRA     | **0.9424** | **0.9413** | **0.9417** | **0.9438** | **0.9471** |
|-----|------|-----------|------------|------------|------------|------------|------------|
| #2  | RMSE | FBP       | 370.0879   | 369.3439   | 427.2159   | 428.6925   | 443.2480   |
|     |      | TVM       | 95.5232    | 97.1169    | 107.0619   | 102.6288   | 104.0917   |
|     |      | FBPConvNet| 30.7892    | 32.0773    | 31.4825    | 29.2609    | 26.8738    |
|     |      | ULTRA     | 29.8473    | 31.0860    | 30.6774    | 28.4754    | 25.8852    |
|     | SSIM | FBP       | 0.3365     | 0.3405     | 0.2982     | 0.2861     | 0.2648     |
|     |      | TVM       | 0.8682     | 0.8687     | 0.8517     | 0.8537     | 0.8592     |
|     |      | FBPConvNet| 0.9451     | 0.9445     | 0.9439     | 0.9450     | 0.9450     |
|     |      | ULTRA     | **0.9502** | **0.9493** | **0.9488** | **0.9498** | **0.9518** |
| #3  | RMSE | FBP       | 366.2710   | 365.4836   | 423.4253   | 423.3997   | 438.7348   |
|     |      | TVM       | 97.8597    | 103.3858   | 106.6264   | 105.2823   | 102.3752   |
|     |      | FBPConvNet| 32.0126    | 33.3269    | 32.5409    | 30.1882    | 27.7669    |
|     |      | ULTRA     | **31.2451**| **32.5243**| **31.9372**| **29.5843**| **26.8926**|
|     | SSIM | FBP       | 0.3457     | 0.3491     | 0.3075     | 0.2975     | 0.2760     |
|     |      | TVM       | 0.8514     | 0.8459     | 0.8346     | 0.8402     | 0.8405     |
|     |      | FBPConvNet| 0.9258     | 0.9248     | 0.9254     | 0.9287     | 0.9311     |
|     |      | ULTRA     | **0.9303** | **0.9286** | **0.9287** | **0.9321** | **0.9368** |

*B.2 Preclinical Mouse*

A dead mouse was scanned by a MARS spectral CT system including one micro x-ray source and one flat-panel PCD, as shown in **Fig. 7**. The used flat-panel PCD consist 660×124 pixels and each of them covers an area of $0.11 \times 0.11$ mm$^2$. The emitting x-ray spectrum with 120kVp is divided into five energy bins: [7.0 32.0], [32.1 43], [43.1 54], [54.1 70] and [70 120]. The distances between the source to the PCD and object are 310 mm and 210 mm, respectively. In this study, we have collected 5760 views with a translation distance set as 88.32mm for helical scanning. Here, the number of views for one circle is 720. The volumetric resolution of reconstructed channels is $0.1 \times 0.1 \times 0.1$ mm$^3$. All reconstructed images were performed by general filtered back-projection method and then they were prepared for testing the networks.

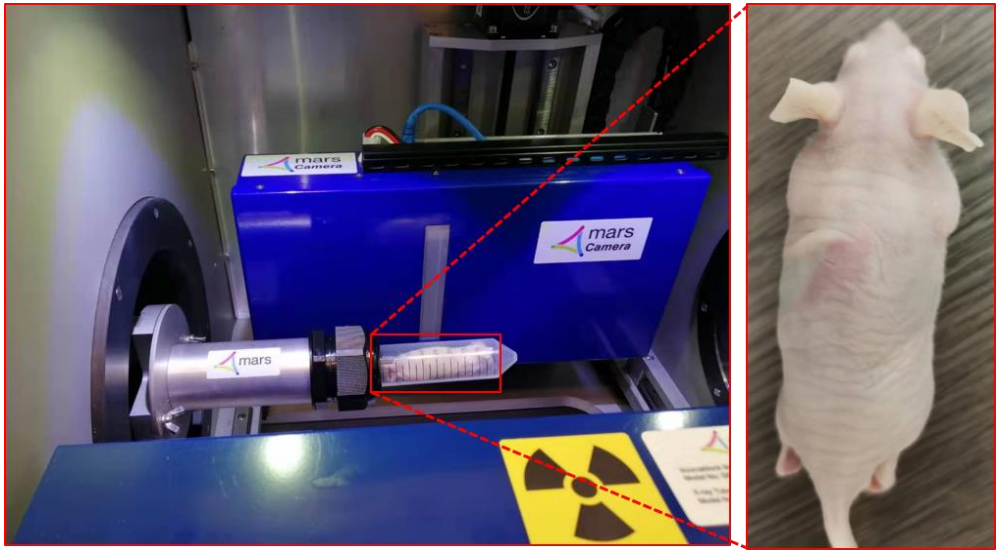

Fig. 7 The set-up MARS scanning system for a preclinical dead mouse.

To demonstrate the advantages of proposed ULTRA method, **Figs. 8** and **9** shows the reconstructed images of two representative cases (4 and 5). The 1st row of images in **Figs. 8** and **9** were reconstructed by the FBP, it could be seen that they are corrupted by severe noises and result in most of details and fine structures being obscured. The image quality could be improved by TVM (2nd row in **Figs. 8** and **9**), it was still difficult to discriminate fine structures as well as image edges. Besides, there were blocky artifacts from TVM results. Compared with these conventional methods, deep learning-based methods could improve the image quality significantly, especially in image edge preservation and feature recovery. Compared with the FBPConvNet results, our proposed method could provide high quality reconstructed image with clear edge and soft tissue structures.

Here, two ROIs were extracted to demonstrate the aforementioned advantages of the ULTRA method and their magnified version are also given in **Figs. 8** and **9**. It could be seen that the profile of soft tissue were blurred in TVM results. Compared with FBPConvNet, the image edge indicated by the arrow **"1"** was blurred, and clear delineation of soft tissue was not well visualized. However, in the ULTRA results, one can easily see the sharp image structure edge. Similarly, the image structures highlighted by the elliptical regions marked with **"2"** and **"3"** further validate the advantages of the ULTRA method. The image structures indicated by arrows **"4"** and **"5"** were more accurate and clearer than those obtained by FBPConvNet. Besides, the extracted elliptical regions marked with **"6"** and **"7"** fully demonstrated our proposed method could obtain higher quality reconstructed images for all energy bins than other methods, including FBPConvNet.

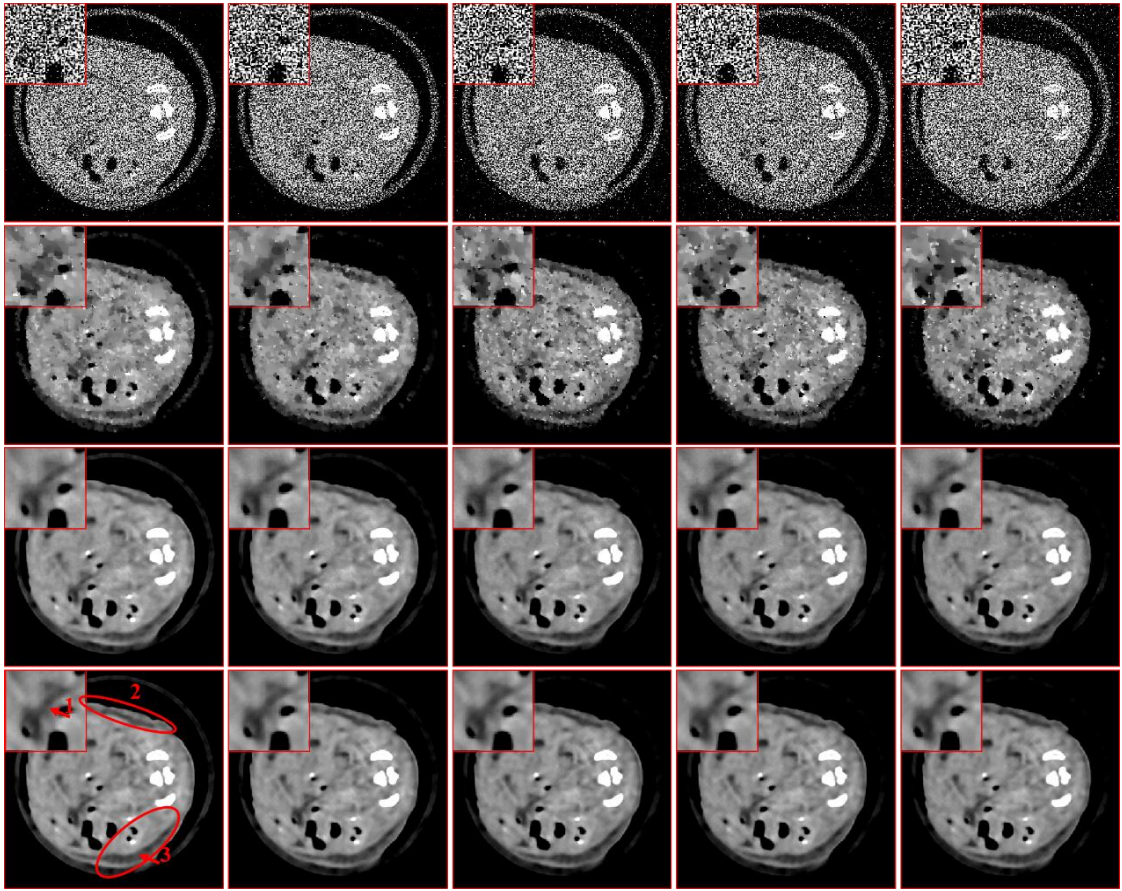

Fig. 8 The reconstruction results of case 1. The 1st-4th rows FBP results with noisy projection, TVM, FBPConvNet and the proposed network. The 1st-5th columns represent reconstructed results from 1st-5th energy bins. The display window is [-250 200] HU.

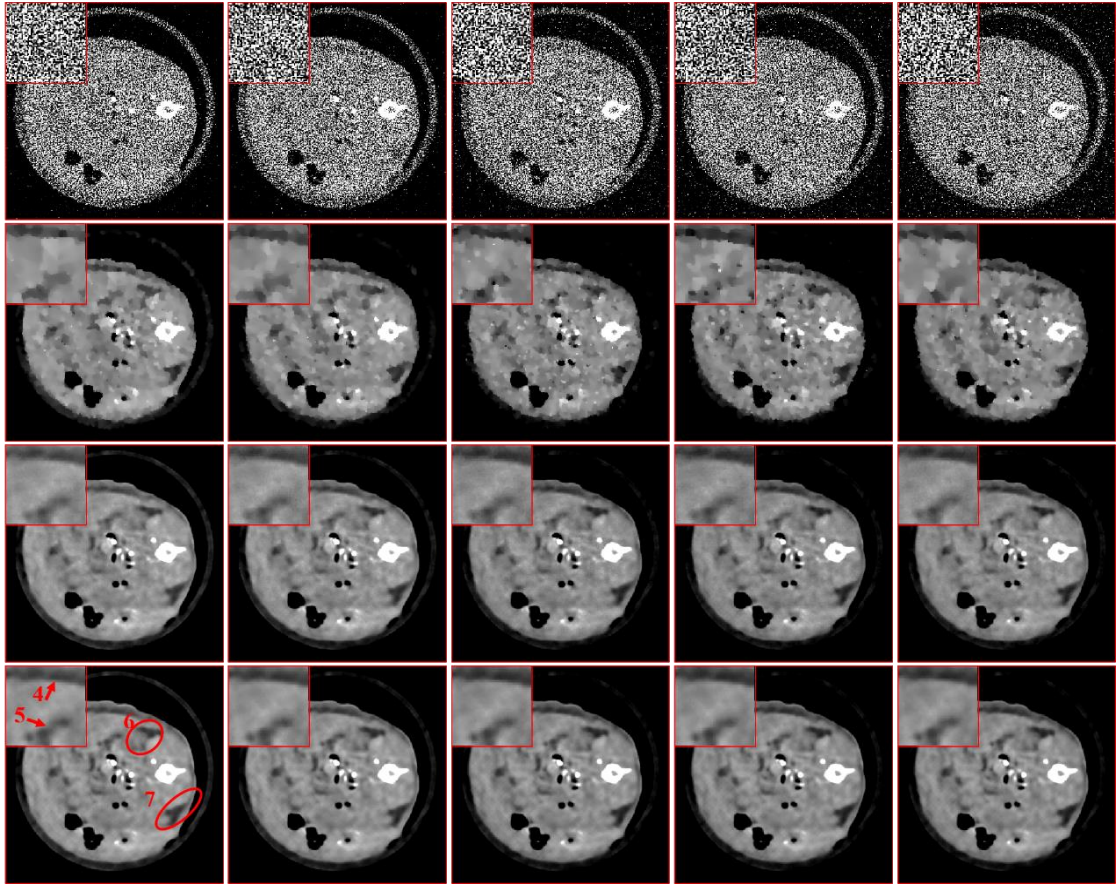

Fig. 9 Same as Fig. 8 but from case 5.

*B.3 Physical Phantom*

To further compare the performance of all reconstruction methods, a home-made physical phantom is scanned by a MARS spectral CT system, as shown in **Fig. 10**. This phantom consists of 14 vials. In order to make different concentrations for different contrast agents, one common iodine contrast agent (Iopamiro 370) and the other gadolinium contrast agent (DOTAREM) are used in this study. Specifically, six vials full with different concentration of iodine, i.e., 8 mg/ml, 4mg/ml, 2mg/ml, 1mg/ml, 0.5mg/ml and 0.25mg/ml. Other six vials full with different concentration of gadolinium (i.e., 8 mg/ml, 4mg/ml, 2mg/ml, 1mg/ml, 0.5mg/ml and 0.25mg/ml) and the rest two vials are water. The 120kVp x-ray source energy spectrum is divided into five energy bins: [7.0, 32.0], [32.1 43], [43.1 54], [54.1 65] and [65.1 120]. The distances starting from the source to the PCD and object are 262 mm and 210 mm, respectively. In this study, we collected 2240 views with the translation as 29.38mm for helical scanning. Here, the number of views for one circle is 720. The volumetric resolution of reconstructed image was $0.1 \times 0.1 \times 0.1$ mm$^3$. All reconstructed images were performed by the general filtered back-projection method and then are prepared for testing the networks.

**Figure 11** shows the reconstructed images of one representative case from the physical phantom. The 1st row of images in **Fig. 11** are reconstructed by the FBP, their results from all energy bins are corrupted by severe noises and lead to the inability to observe the image edge and structures. TVM (2nd row in **Fig. 11**) could discriminate overall image structures. However, the profiles of the vials were blurred with blocky artifacts. Compared with these methods, deep learning-based could improve the image quality a lot in the aspect of clear image edge preservation. Because the structure of this physical phantom was relatively simple, both FBPConvNet and our proposed ULTRA method could provide similar high quality reconstructed spectral CT images.

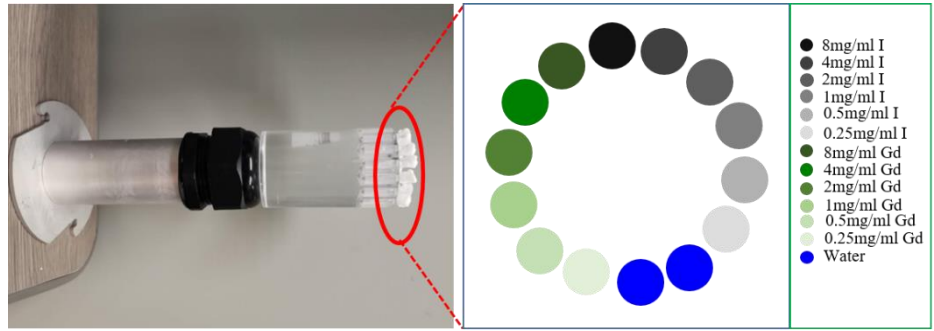

Fig. 10 The set-up MARS scanning system for a home-made phantom.

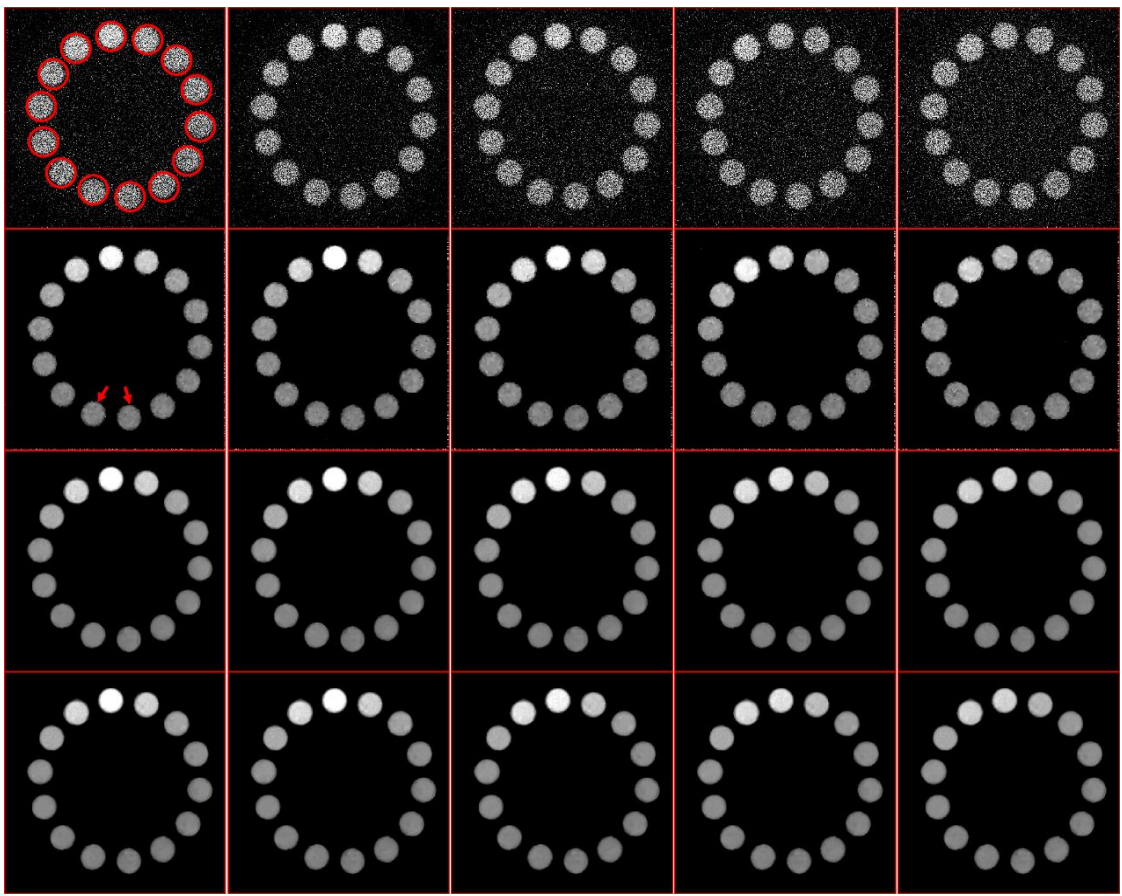

Fig. 11 The reconstruction results of case 6. The 1st-4th rows FBP results with noisy projection, TVM, FBPConvNet and the proposed network. The 1st-5th columns represent reconstructed results from 1st-5th energy bins. The display window is [-400 400] HU.

The great advantage of spectral CT is that can provide accurate material decomposition results, which is difficult to be provided by traditional CT. The accurate material decomposition results are beneficial to clinical diagnosis (Mukundan Jr, et al., 2006; Yeh, et al., 2017). To quantitatively compare all reconstruction results in terms of material decomposition, the results in case 6 reconstructed using different algorithms are decomposed into three materials, i.e., water, iodine, and gadolinium. In this study, the direct inversion (Wu, Chen, Vardhanabhuti, Wu, & Yu, 2019) was employed. **Figure 12** shows the material decomposition results. It can be seen that FBP yielded so large errors that we could not discriminate iodine from gadolinium components. TVM improved the material decomposition accuracy but the associated outlier artifacts appeared in iodine and gadolinium regions. Compared with FBPConvNet results, our proposed ULTRA method provided higher image quality with improved material maps, especially for the vials indicated by arrows "1" and "2". This is because our proposed ULTRA provided high-quality reconstruction results in the first place, directly facilitating the material decomposition."

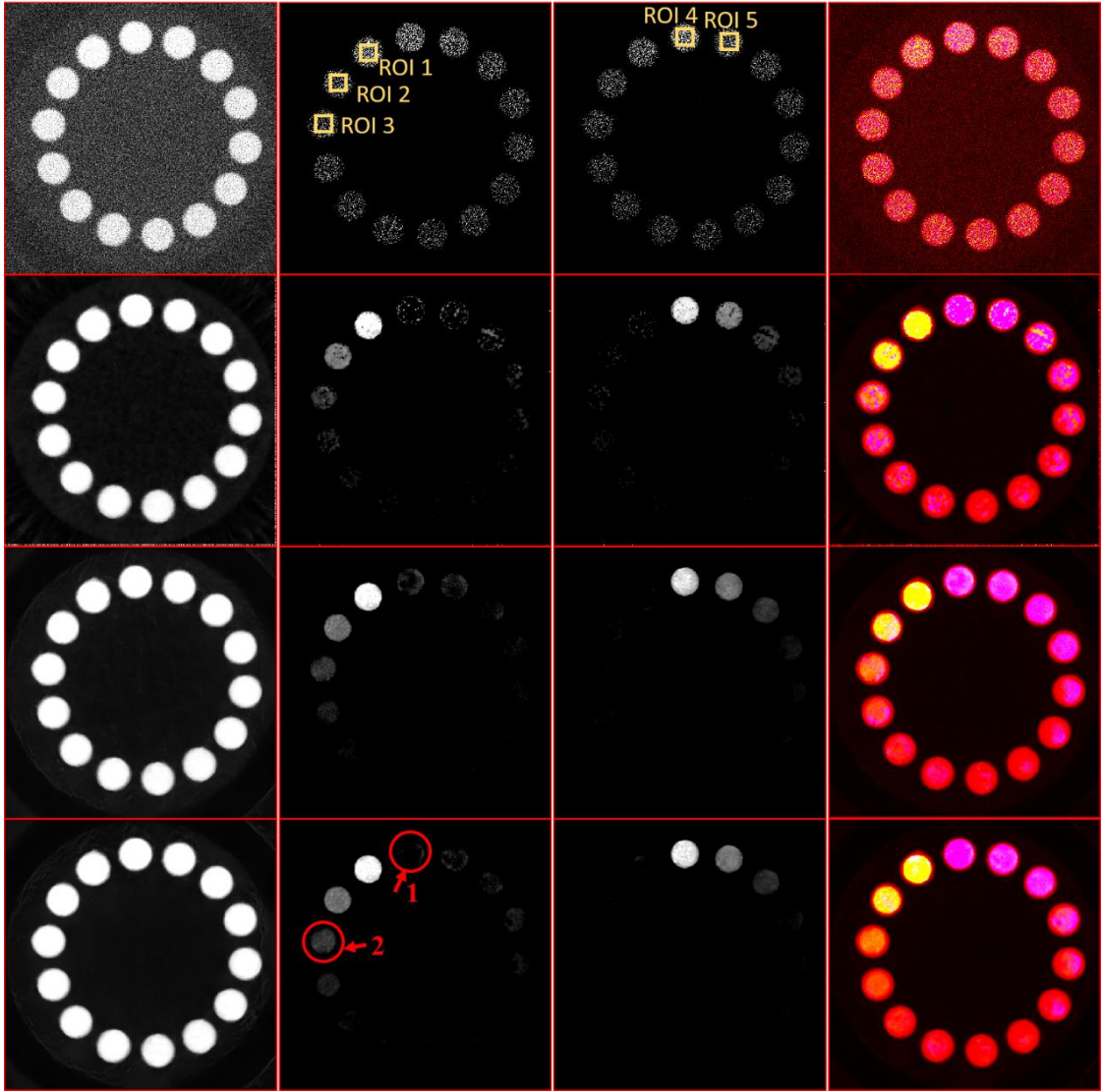

Fig. 12 The material decomposition results of case 6. The 1$^{st}$-4$^{th}$ rows FBP results with noisy projection, TVM, FBPConvNet and ULTRA. The 1$^{st}$-3$^{rd}$ columns represent water, gadolinium contrast agent and iodine contrast agent, where their display windows are [0 1]g/ml, [0.3 8]mg/ml and [0.3 8]mg/ml. 4$^{th}$ column represents color rending image, where yellow, magenta and red represent gadolinium, iodine and water respectively.

To further quantitatively compare material decomposition results, the RMSE and peak signal-noise ratio (PSNR) are employed in this study. In addition, we introduce a new metric to compute the relative bias (RB) of decomposed materials. RB is a metric index to compute all pixels bias between decomposed materials and the ground truth, which can be defined as

$$\text{RB} = \frac{1}{W_1 \times H_1} \sum_{w_1=1}^{W_1} \sum_{h_1=1}^{H_1} \left\{ \frac{|X_{w_1 h_1} - \text{Ref}_{w_1 h_1}|}{\text{Ref}_{w_1 h_1}} \right\} \times 100\%, \tag{24}$$

where $\mathbf{X}_{w_1 h_1}$ and $\mathbf{Ref}_{w_1 h_1}$ represent $(w_1, h_1)^{\text{th}}$ elements of material image $\mathbf{X}$ and ground truth $\mathbf{Ref}$. From Eq. (24), it can be seen that RB can reflect the relative bias with the ground truth. The smaller of RB is, the greater of material decomposition accuracy. Since the specified contrast agents concentrations of different vials are known in advance, the quantitative results from five extracted ROIs in **Fig.12** are listed in **Table II**. It can be seen that our proposed ULTRA network can always obtain the smallest RMSEs and RBs with the highest PSNR. Compared with TVM and FBPConvNet, our proposed network can reduce RB from 10.08% and 7.64% to 5.17% for ROI 1, 12.55% and 11.49% to 10.51% for ROI2, 30.29% and 16.88% to 12.56% for ROI3, 9.37% and 11.45% to 7.72% for ROI4, 15.83% and 8.01% to 6.03% for ROI5. In terms of ROIs 1-3 from gadolinium contrast agents, RMSEs of our proposed network can reduce to 0.5262 mg/ml, 0.4978 mg/ml and 0.3241 mg/ml respectively. In contrast, ROIs 4-5 from iodine contrast agents, RMSEs of our proposed network can reduce to 0.7301 mg/ml and 0.3087 mg/ml. As for PSNR index, our proposed network can obtain remarkable improvements with other algorithms.

Table II. Quantitative evaluation results of the material decomposition of five extracted ROIs from the physical phantom

| Metrics | Methods | ROI 1 | ROI 2 | ROI 3 | ROI 4 | ROI 5 |
|---|---|---|---|---|---|---|
| RMSE(mg/ml) | FBP | 7.3497 | 4.8297 | 3.9257 | 6.1718 | 4.1137 |
| | TVM | 1.1268 | 0.8254 | 0.8632 | 1.1521 | 0.8547 |
| | FBPConvNet | 0.7376 | 0.5611 | 0.4471 | 1.0559 | 0.3988 |
| | ULTRA | **0.5262** | **0.4978** | **0.3241** | **0.7301** | **0.3087** |
| PSNR(db) | FBP | 30.80 | 34.45 | 36.25 | 32.32 | 35.85 |
| | TVM | 47.09 | 49.80 | 49.51 | 46.90 | 49.50 |
| | FBPConvNet | 50.77 | 53.15 | 55.12 | 47.66 | 56.12 |
| | ULTRA | **53.71** | **54.49** | **57.92** | **50.86** | **58.34** |
| RB (%) | FBP | 82.29 | 103.06 | 144.01 | 68.30 | 90.77 |
| | TVM | 10.08 | 12.55 | 30.29 | 9.37 | 15.83 |
| | FBPConvNet | 7.64 | 11.49 | 16.88 | 11.45 | 8.01 |
| | ULTRA | **5.17** | **10.51** | **12.56** | **7.72** | **6.03** |

## V. Discussions and Conclusion

High-quality spectral CT imaging has great significance for future applications, i.e., feature and details recovery, image structure preservation, material discrimination and decomposition, volume effect removal and radiation dose reduction. As a result, spectral CT can benefit the biological and function evaluation of different tissues. Traditional iterative reconstruction method incorporating prior knowledge model can achieve outstanding performances in some cases, they are greatly limited by computational costs and parameters selections in practice. Again, the unit of PCD is very small (i.e., 0.1×0.1mm² or 0.11×0.11mm²), which is a degree of magnitude smaller than the usual 1×1mm² in conventional medical CT. Besides, PCD collects the projections from several energy bins, such as 5 or 8. Therefore, the storage of spectral CT projection with helical geometry may be hundreds of times bigger than that obtained by conventional helical CT for a single patient. In such a case, if the prior knowledge based iteration methods is employed, the computational costs may be unacceptable and even unfeasible. However, if only the analytical method is used for photon-counting spectral CT, the noise within projections will obscure most of image details and features rendering image quality uninterpretable in clinical practice. To address these issues, deep learning based spectral CT imaging provided a good solution. In this study, we developed a deep learning based reconstruction method for spectral CT.

Our methods result in reduced image noise while maintaining image details. Compared with FBPConvNet designed for conventional CT image reconstruction, the advantages of our ULTRA are in the following three aspects. First, while the existing U-net based methods do not fully encode image features and details since image structures are not always shared among downsampling layers. To overcome this problem, we have included multiple skip connections between different downsampling layers to realize multi-scale feature fusion and multichannel filtration, which help improve image quality. Second, to avoid image blurring due to the $L_2^2$ loss, we have proposed a general $L_p^p$- loss to emphasize the sparsity of the output. Third, because different energy channel images come from the same object, spectral CT images in these channels share similar structures and features. To incorporate such prior into the network, energy channel-wise images were converted into Hounsfield unit independently, and anisotropic total variation was incorporated into the network as an additional loss term.

Although our network has produced excellent results, there is still potential for future improvements. First, a more accurate and more diverse database should be constructed to train the network better. In this study, we numerically synthesized a database but it is not perfect since many practical factors, such as pile-up and charge sharing, are not effectively modeled. The simulation and preclinical cases should be better matched in our future work, and fine-tuned with real scans. Second, our proposed network still sometimes messes up subtle structures. This could be because some image details and features are missing during downsampling process. To overcome this problem, the generative adversarial network (GAN) may be used (Radford, Metz, & Chintala, 2015). ). Third, because channel-wise spectral CT images come from the same object, they share similar structures and features. Compared with 2D neural networks, 3D neural networks may be beneficial to enhance information sharing and feature fusion between energy channels. However, the network training and computational costs for 3D ULTRA will much larger than 2D ULTRA. In this study, we have focused on 2D ULTRA to establish the feasibility and merits. In a follow-up study, we plan to develop efficient techniques and handle 3D images directly in different energy channels for even better imaging performance. Fourth, the commercial spectral CT system works in spiral cone-beam geometry, how to extend our proposed network for spiral cone-beam spectral CT is an important topic for future investigation. Fifth, since many existing deep reconstruction networks are subject to instabilities (Antun, Renna, Poon, Adcock, & Hansen, 2020), our feed-forward ULTRA network could be also subject to adversarial attacks. As a follow-up project, we plan to stabilize ULTRA in our recently proposed analytic mapping, compressive sensing, iterative reconstruction, and deep learning (ACID) framework (Wu, Hu, Wang, et al., 2020).

In conclusion, we have established a Skip-encode U-net with $L_p^p$-loss and Anisotropy Total-variation (ULTRA) training framework for spectral CT imaging. Aided by training data, our approach learns complex structured features more effectively. At a relatively low computational cost, the proposed network can achieve significant image quality improvement. In general, the proposed ULTRA can generate promising results with anatomical structure preservation and noise suppression. Better visual evaluation results are obtained from numerical simulation, preclinical mice and physical phantom. Besides, the material decomposition results from the physical phantom confirm that our proposed ULTRA networks outperform other competitors contributing much advancementfor spectral CT imaging.

**Acknowledgement:** This study is partially supported by the Li Ka Shing Medical Foundation.